\documentclass[reprint,aps,nofootinbib,superscriptaddress, prl, nobibnotes]{revtex4-2}

\usepackage{braket}
\usepackage{slashed}
\usepackage{graphicx}
\usepackage{xcolor}
\usepackage{enumitem}
\usepackage{siunitx}
\usepackage{mathtools}
\usepackage{dsfont}
\usepackage[hidelinks]{hyperref}
\usepackage{amsmath,amsthm,amsfonts,amscd,amssymb} 
\usepackage[justification=Justified,
   format=plain]{caption}
\usepackage{subcaption}

\newcommand{\re}{\mathrm{Re}\,}
\newcommand{\im}{\mathrm{Im}\,}

\newcommand{\gas}{\mathrm{gas}}
\newcommand{\sph}{\mathrm{sph}}

\newcommand{\mb}[1]{\mathbf{#1}}

\begin{document}

\title{Three-dimensional squeezing of optically levitated nanospheres}

\author{Giacomo Marocco}
\thanks{gmarocco@lbl.gov}
\affiliation{Physics Division, Lawrence Berkeley National Laboratory, Berkeley, California 94720, USA}
\author{David C. Moore}
\thanks{david.c.moore@yale.edu}
\affiliation{Wright Laboratory, Department of Physics, Yale University, New Haven, Connecticut 06511, USA}
\author{Daniel Carney}
\thanks{carney@lbl.gov}
\affiliation{Physics Division, Lawrence Berkeley National Laboratory, Berkeley, California 94720, USA}

\begin{abstract}
We propose a protocol to measure impulses beyond the standard quantum limit. The protocol reduces noise in all three spatial dimensions and consists of squeezing a mechanical system's state via a series of jumps in the frequency of the harmonic potential. We quantify how decoherence in a realistic system of an optically levitated, dielectric nanoparticle limits the ultimate sensitivity. We predict that $\sim$10 dB of squeezing is achievable with current technology, enabling quantum-enhanced detection of weak impulses. 
\end{abstract}

\maketitle

Mechanical quantum sensing of weak forces and impulses is playing an increasingly important role in both standard lab metrology~\cite{Aspelmeyer:2013lha, Gonzalez-Ballestero:2021gnu} as well as fundamental physics~\cite{moore2021searching, Beckey:2023shi}. Notable use cases include pressure sensing~\cite{Barker:2023gqa}, gravimetry~\cite{Schmole:2016mde, griggs2017sensitive, westphal2021measurement, Fuchs:2023ajk}, gravitational wave detection~\cite{bersanetti2021advanced, abe2022current, ganapathy2023broadband, Aggarwal:2020umq, Carney:2024zzk, Jungkind:2025oqm}, tests of both classical~\cite{Geraci:2010ft} and quantum gravity~\cite{romero2011quantum, bose2017spin, marletto2017gravitationally}, and searches for dark matter~\cite{Carney:2019pza, Carney:2019cio, Monteiro:2020wcb, Afek:2021vjy, Higgins:2023gwq, Day:2023mkb, Gosling:2023lgh, Amaral:2024rbj, Qin:2025jun, Tseng:2025rlo} and other new particles~\cite{Moore:2014yba, Rider:2016xaq, Carney:2022pku}. 

For many of these applications, the essential problem is to sense an approximately instantaneous force $\mathbf{F}(t) = \Delta p \, \delta(t) \hat{\mathbf{n}}$, such as that imparted by a particle colliding with the sensing element. A key goal is to sense the smallest possible impulse $\Delta p$. This detection problem has a standard quantum limit  $\Delta p_\mathrm{SQL}=\sqrt{m \omega}$~\cite{clerk2004quantum}, which sets a floor for the smallest detectable impulse above quantum vacuum noise in terms of the sensing element's mass $m$ and trapping frequency $\omega$, and is close to the sensitivity of current devices~\cite{magriniRealtime2021, tebbenjohanns2021quantum, Tseng:2025rlo}. Many methods to detect impulses below the SQL exist, using squeezed readout light~\cite{caves1981quantum, kimble2001conversion, schnabel2017squeezed, mason2019continuous, Lee:2025hkp}, squeezing of the mechanical state itself~\cite{rossiQuantum2024, wuSqueezing2024, kamba2025quantum,Skrabulis:2026kun}, and/or backaction evasion~\cite{thorne1978quantum, braginsky1990gravitational, khalili1996speed, kimble2001conversion, ghosh2020backaction, Richman:2023mak}. However, to date, these protocols have been limited to one spatial dimension. For many signals of interest, which may act along an \textit{a priori} unknown direction $\hat{\mathbf{n}}$, this is a major hindrance to practical use.

In this paper, we propose a fully three-dimensional squeezing protocol tailored to levitated mechanical sensors. The experimental requirements are minimal; a single standard trapping and readout laser is sufficient. The core idea is a straightforward generalization of existing one-dimensional protocols based on modulating the trap frequency. As a case study, we analyze levitated dielectric beads with experimental parameters corresponding to existing experimental implementations, and find that standard noise and decoherence mechanisms should still enable at least a decade of noise reduction below the SQL.

\begin{figure*}[t!]
    \centering
    \includegraphics[width=\linewidth]{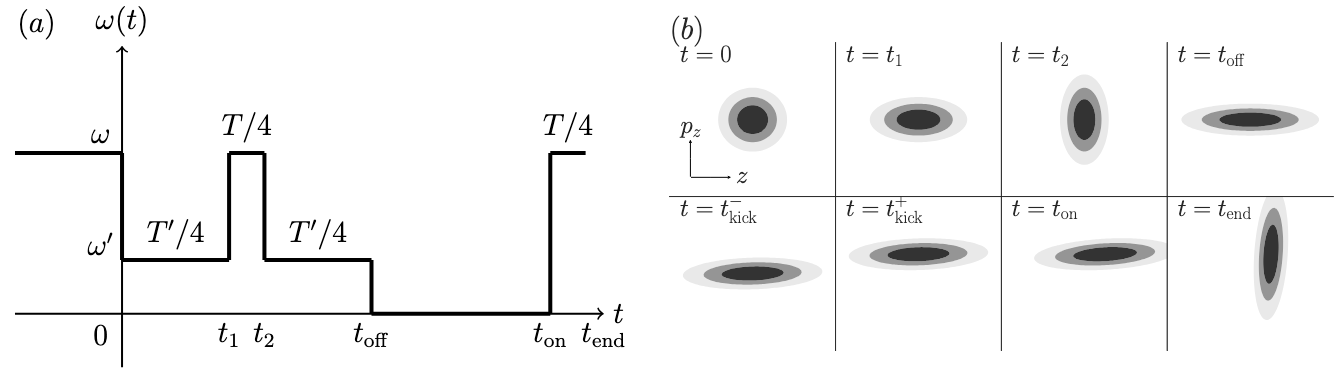}
    \caption{(a) An example of the optical potential as a function of time for the squeezing protocol. The particle is cooled close to the ground state of the trap at frequency $\omega$, and at $t=0$ the squeezing begins. After two cycles of squeezing (i.e. two time segments in the smaller potential $\omega'$), the trap is switched off at $t=t_\mathrm{off}$. The particle evolves freely until $t=t_\mathrm{on}$, at which point the a phase rotation begins. At $t=t_\mathrm{end}$, the particle's position is measured. (b) Projection of the particle's Wigner function at different times when an impulse acts on the particle at $t = t_\mathrm{kick}$. We sketch the Wigner functions just before the kick (at $t = t^-_\mathrm{kick}$) as well as just after (at $t = t^+_\mathrm{kick}$). 
    }
    \label{fig:protocol}
\end{figure*}

\textit{Squeezing protocol.}--- The protocol we study involves the sequential squeezing of the motional state of the nanoparticle by modulating the harmonic potential induced by the optical tweezer, as demonstrated recently in one dimension~\cite{rossiQuantum2024, kamba2025quantum,Skrabulis:2026kun}. The general problem of trying to engineer the potential landscape $V(\mb{x}, t)$ to steer from one state to another in a time-optimal way is well understood: in this conservative setting, ``bang-bang"  protocols\footnote{A bang-bang control protocol is one in which $V(\mb{x}, t)$ alternates between two possible potential landscapes.} are time-optimal~\cite{salamon2009maximum, chen2010fast, stefanatos2010frictionless}. 

The basic protocol is illustrated in Fig.~\ref{fig:protocol}. The stiffness of the harmonic potential is changed quasi-instantaneously between two values. By judicious choice of the time spent at each trap frequency, the evolution operator $U(t) = \exp\{ - i \int_0^t dt' V(t')\} $ dynamically generates squeezing~\cite{janszky1992strong, graham1987squeezing}.   

The translational modes of the nanoparticle are described by position operators $x_i = (x,y,z)$ and conjugate momenta $p_i = (p_x, p_y, p_z)$, with $z$ parallel to the laser propagation direction. The time-dependent potential is of the form
$    V = m \omega^2_{ij}(t) x^i x^j/2$, where $\omega^2_{ij}$ is a diagonal matrix
$    \omega^2_{ij}(t) = \mathrm{diag}\big(
        \omega_x^2(t),
        \omega_y^2(t),\omega_z^2(t)\big)
$ whose elements each take two possible values $\omega_i>\omega'_i$. In an optical tweezer, this frequency switching is typically implemented by tuning the laser power, since $\omega_i(t) \propto \sqrt{P_\mathrm{L}}$. Each mode begins in potential $\omega_i$ and is cooled such that it sits close to the ground state. Upon changing the harmonic frequencies to $\omega'_i$ for a quarter period or three-quarter period, the variance in $p_i$ is decreased by a factor $(\omega'_i/\omega_i)^2$ (see End Matter), i.e., the time spent in  $\omega'_i$ must satisfy
$t_1 = (2k'_i+1) \pi/2\omega'_i$, with $k'_i$ integers. We may rewrite this latter condition as a constraint on the ratio of the frequencies
\begin{align}
    \frac{\omega'_{x}}{2k'_x + 1}=\frac{\omega'_{y}}{2k'_y + 1}=\frac{\omega'_{z}}{2k'_z + 1},
    \label{eqn:harmonicCondition}
\end{align}
which must be satisfied in order for all three directions to be simultaneously momentum-squeezed. Abstractly, the harmonic condition Eq.~\eqref{eqn:harmonicCondition} of course has many solutions, and we now show they are practically attainable.

For a dielectric nanoparticle whose confining potential is provided by an optical tweezer, the ratio of frequencies along different axes is primarily set by the numerical aperture (NA) of the lens and the polarization of laser light~\cite{novotny2012principles}. With linearly polarized light, one may tune NA such that any two of $\omega'_i$ satisfy \eqref{eqn:harmonicCondition}, but the third mode does not. On the other hand, with circularly polarized light, the two transverse modes are automatically degenerate $\omega^{(\prime)}_x = \omega^{(\prime)}_y = \omega^{(\prime)}_\perp$. Furthermore, we show in the End Matter that a Gaussian beam incident on a lens with NA~$\approx 0.85$ solves the harmonic condition with $\omega'_\perp = 3 \omega'_z$ (see Fig.~\ref{fig:frequencyRatio}), indicating that the all three modes will be simultaneous momentum-squeezed.

One may further squeeze the nanoparticle by stiffening the harmonic potential back to $\omega_i$ for a time that satisfies $t_2 = (2k_i+1)\pi/2\omega_i$. Since $\omega_i(t) \propto \sqrt{P_\mathrm{L}}$, if the ratios of trap frequencies satisfy \eqref{eqn:harmonicCondition} at one laser power, they will satisfy the equivalent harmonic condition at all laser powers. This implies that setting $t_2 = \pi/2\omega_z$ is sufficient to ensure the harmonic condition $t_2 = 3\pi/2\omega_\perp$ is satisfied along the other two axes. After a further time $t_1$ in the less stiff $\omega'_i$ potential, the $p_i$ variance is again a factor $(\omega'_i/\omega_i)^2$ smaller, in the absence of decoherence. In total, after $n_\mathrm{cycles}$ of such cycles, the variance of the 3D vector $p_i$ is multiplicatively decreased by $(\omega'_i/\omega_i)^{2n_\mathrm{cycles}}$ in the idealized case.

\textit{Decoherence.}--- In a realistic implementation of the above squeezing protocols, there are a number of sources of decoherence whose impact we now take into account. The dynamics of the nanoparticle's density matrix $\rho$ obeys a stochastic difference equation~\cite{jacobs2006straightforward, romero2011optically,gonzalez2019theory}
\begin{align}
    d\rho = -i[H_0, \rho] dt - \sum_{i=1}^3 \big( \mathcal{D}_{x_i}[\rho] dt +  \mathcal{H}_{x_i}[\rho] dW \big),
    \label{eqn:SME}
\end{align}
where $H_0 = \sum_i p_i^2/2m + V$, $\mathcal{D}_{x_i}[\rho] = \Gamma_{\mathrm{tot},i}[x_i,[x_i,\rho]]/2x_{0,i}^2$
is the dissipator associated with position measurements, and 
$\mathcal{H}_{x_i}[\rho] = \sqrt{\eta_i \Gamma_i}\big(\{x_i,\rho\} - 2 \langle x_i \rangle \rho\big)/x_{0,i}$
describes the stochastic jumps in the state conditioned on the measurement outcomes\footnote{We note that here we ignore the effect of momentum damping of the oscillator, induced for instance by radiation damping~\cite{novotny2017radiation}, since its effects are negligible over a number of periods much smaller than the oscillator's quality factor.
}. Here, $\Gamma_{\mathrm{tot},i}\,(\Gamma_i)$ is the total (measurement-induced) decoherence rate along each axis, $\eta_i$ the respective measurement efficiencies, and $x_{0,i} = 1/\sqrt{m \omega_{i}}$. The stochastic dissipator depends on a Wiener increment $dW$ satisfying $dW^2 = dt$, which encodes the random outcome of the weak, continuous position measurements.

Numerous sources of decoherence contribute to the total decoherence rates $\Gamma_{\mathrm{tot},i}$~\cite{romero2011optically,gonzalez2019theory}. For one, the measurement backaction due to the photon-oscillator scattering localizes the oscillator in position-space. The rate $\Gamma_i$ depends on the intensity of the laser and so is time-dependent in our protocol. Other sources of decoherence, due to gas scattering~\cite{vacchini2000completely, hornberger2003collisional, hornberger2006master, stickler2016spatio}, thermal emission~\cite{hackermuller2004decoherence, Schafer:2024uco}, or to desorption~\cite{Schafer:2025wtf}, are all subdominant given the experimental parameters of Tab.~\ref{tab:params} (see End Matter).

We assume that the oscillator is initially in a Gaussian state, in particular the ground state. Since the potentials and dissipators we consider are at most quadratic in the oscillator's position coordinates, the state remains Gaussian throughout the protocol. As such, the state is fully characterized by its first two cumulants, i.e., the expectation values of position $\langle x_i \rangle$ and momentum $\langle p_i \rangle $ as well as the variances $V_{x_ix_i} := \langle x_i^2 \rangle - \langle x_i \rangle^2 $, $V_{p_ip_i} := \langle p_i^2 \rangle - \langle p_i \rangle^2 $ and covariance $C_{x_ip_i} := \langle \{x_i, p_i\}\rangle/2 - \langle x_i \rangle \langle p_i \rangle $. The evolution of these quantities may then be solved to understand the dynamics of decoherence and squeezing.

\begin{figure}[t!]
    \centering
    \includegraphics[width=\linewidth]{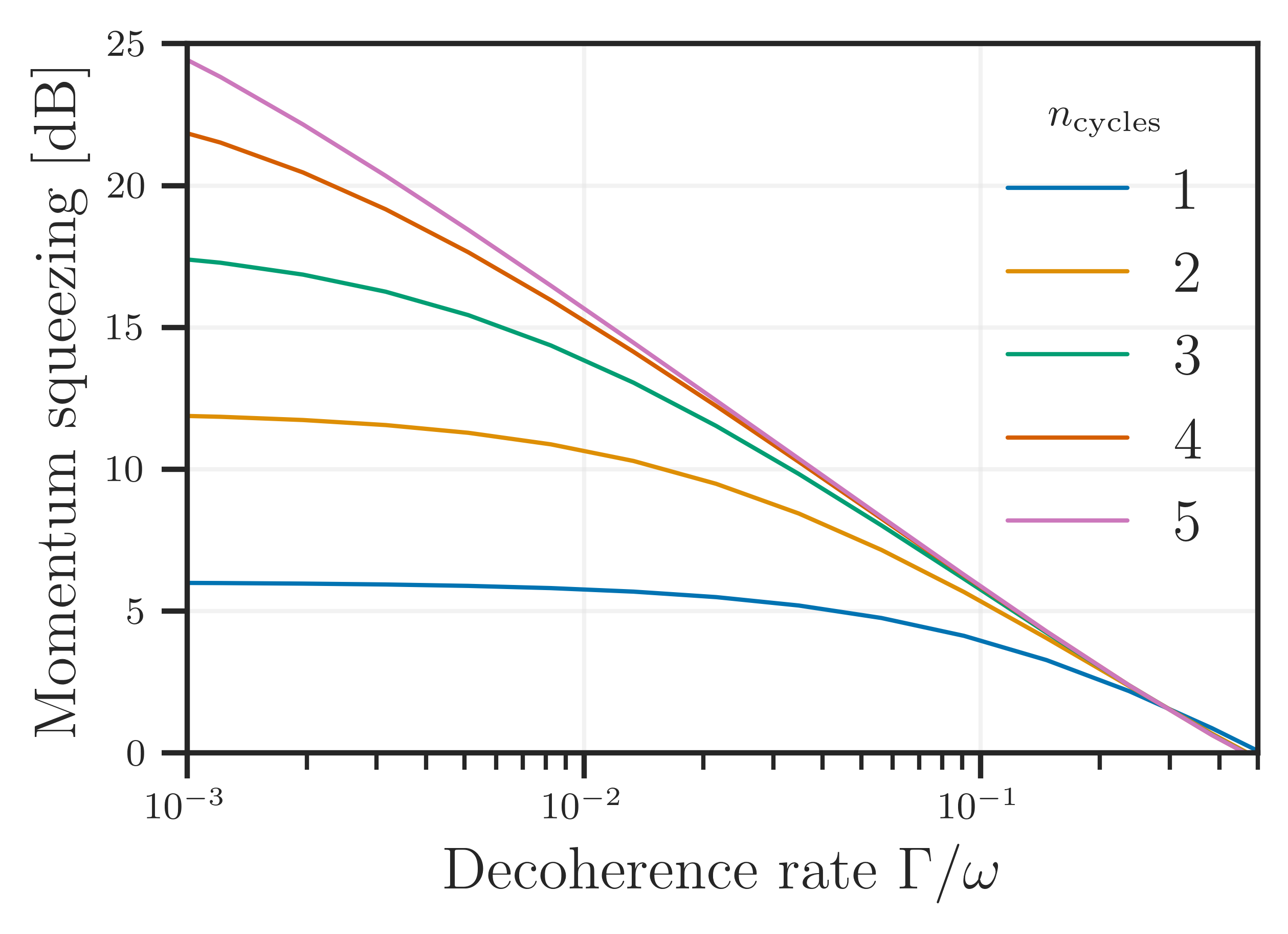}
    \caption{The squeezing of the momentum variance, measured in decibels, as a function of the decoherence rate. We plot the amount of squeezing for a different number of squeezing cycles $n_\mathrm{cycles}$ for measurement efficiency $\eta = 0.2$ and $\omega'/\omega=0.5$. }
\label{fig:squeezing_vs_decoherence}
\end{figure}

The deterministic equations of motion for the covariances are~\cite{doherty1999feedback}
\begin{align}
    \begin{split}
        \dot{V}_{x_ix_i} &= 2\frac{ C_{x_ip_i}}{m} - 4 \eta_i \frac{\Gamma_i(t)}{x_{i,0}^2}V_{x_ix_i}^2, \\ 
        \dot{V}_{p_ip_i} &= -2 m \omega_i^2(t)C_{x_ip_i} + \frac{ \Gamma_i(t)}{x_{i,0}^2} - 4 \eta_i \frac{\Gamma_i(t)}{x_{i,0}^2}C_{x_ip_i}^2, \\
        \dot{C}_{x_ip_i} &= \frac{V_{p_ip_i}}{m} - m \omega_i^2(t) V_{x_ix_i} - 4 \eta_i \frac{\Gamma_i(t)}{x_{i,0}^2}V_{x_ix_i}C_{x_ip_i}.
    \end{split}
    \label{eqn:cumulants}
\end{align}
In the absence of the non-linearities induced by conditioning the state on the measurement outcomes, i.e., with $\eta_i=0$, the particle behaves as a harmonic oscillator subject to the random diffusion from scattering of the laser photons. In this case, we derive an analytic solution to \eqref{eqn:cumulants}, given in the the End Matter. The asymptotic momentum variance in the limit of a large number of squeezing cycles, starting from the ground state with $V_{p_ip_i} = m\omega_i/2$,  is
\begin{align}
    \lim_{N_\mathrm{cycles} \to \infty} V_{p_ip_i} = \frac{m\omega_i}{2} \times \frac{\pi}{2} \frac{\Gamma_i}{\omega_i} \frac{\omega_i^2 + \omega'^2_i}{\omega_i^2 - \omega'^2_i},
    \label{eqn:asymptoticSqueezing}
\end{align}
which demonstrates that the maximal amount of squeezing is largely determined by the dimensionless decoherence rate $\Gamma_i/\omega_i$. As long as $\omega_i\gg \omega'_i$, the last factor tends to one and the dependence on $\omega'_i/\omega_i$ is soft. In the presence of continuous position measurements,  the equations of motion Eq.~\eqref{eqn:cumulants} are non-linear and we numerically solve them, as shown in Fig.~\ref{fig:squeezing_vs_decoherence}. We see that the squeezing quickly tends to the asymptotic Eq.~\eqref{eqn:asymptoticSqueezing}, even for a small number of squeezing cycles. For small decoherence parameters, the amount of squeezing is log$_{10}(2 \omega/\omega')^{n_\mathrm{cycles}} \,\mathrm{dB} \approx 
    n_\mathrm{cycles}  \times 6.9 $ dB.

\begin{figure*}[t!]
    \centering
    \includegraphics[width=\linewidth]{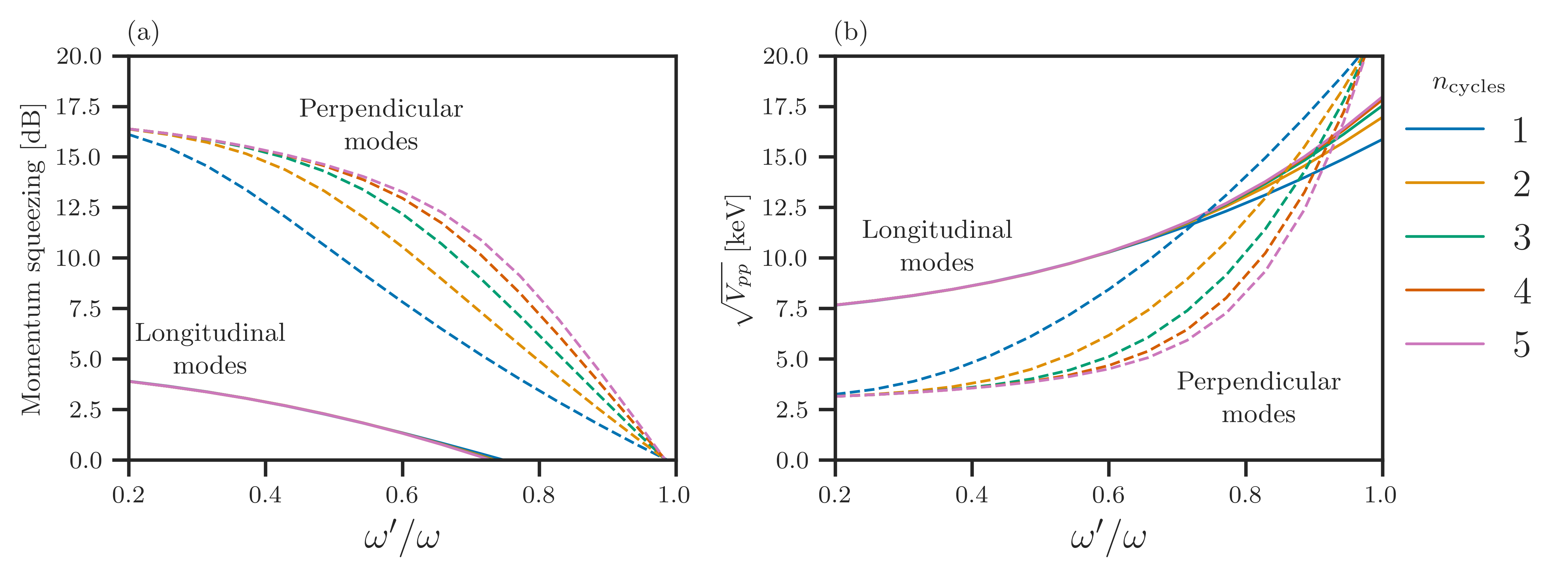}
    \caption{(a) The squeezing of the momentum variance as a function of the frequency ratio $\omega'/\omega$ for a varying number of squeezing cycles $n_\mathrm{cycles}$. We distinguish between longitudinal modes (parallel to the laser direction) and those that are perpendicular due to their different decoherence rates. We note that the lines for the longitudinal mode lines lie largely on top of one another since the squeezing quickly reaches its asymptotic value. (b) The width of the state in momentum space in absolute units, as a function of the frequency ratio.}
\label{fig:squeezing_vs_frequency}
\end{figure*}

\begin{table}
\begin{ruledtabular}
\begin{tabular}{lll  } 

 sphere radius & $R_\mathrm{sph}$ & \SI{80}{nm} \\
 laser wavelength & $\lambda$ &  \SI{1550}{nm}  \\
 longitudinal mechanical frequency & $\omega_z/2\pi$ & \SI{50}{kHz} \\
 relative dielectric constant~\cite{malitson1965interspecimen, romero2011quantum} & $\epsilon_\mathrm{r}$ & 2.07 + $10^{-10}i$\\
 density & $\rho$ & 
 \SI{2.0}{g/cm^3} \\
 blackbody absorption fraction~\cite{chang2010cavity} & $\mathrm{Im}  \tfrac{\epsilon_\mathrm{bb} - 1}{\epsilon_\mathrm{bb} + 2}$ & 0.1 \\
 gas pressure & $P$ & \SI{e-10}{mbar} \\
 gas temperature & $T$ &
\SI{300}{K} \\
\end{tabular}
 \caption{The experimental parameters considered, closely following those demonstrated in~\cite{Tseng:2025rlo}.}
 \label{tab:params}
 \end{ruledtabular}
 \end{table}

For a dielectric nanosphere with the nominal parameters of Tab.~\ref{tab:params}, we calculate the decoherence rates to be approximately $\Gamma_{z} \approx 0.25~\omega_z$ and $\Gamma_x = \Gamma_y \approx 4.5\times 10^{-3}~\omega_\perp$, which are dominated by laser recoil heating. Fixing these decoherence rates, we plot in Fig.~\ref{fig:squeezing_vs_frequency} the amount of squeezing we may achieve as a function of the ratio of the two trap frequencies. Since the modes perpendicular to the laser propagation experience a stiffer potential and less decoherence [see Eq.~\eqref{eqn:heatingRate2}], these modes become much more squeezed than the longitudinal mode, with more than 15 dB of squeezing predicted. In absolute units, this implies that upon squeezing the perpendicular modes are actually more sensitive to impulses than the longitudinal mode, even though the situation is reversed when operating at the SQL. We again find good agreement with the analytic scaling estimate of \eqref{eqn:asymptoticSqueezing}.

The time taken for the squeezing protocol is determined by the number of cycles and by the mechanical frequencies of the two harmonic potentials $\omega_{z,0}$ and $\omega_{z,1}$. We see in Fig.~\ref{fig:squeezing_vs_frequency} that the degree of squeezing is maximal for small $\omega'/\omega$, and almost saturated after one cycle for small frequency ratios. Hence the total squeezing time is given by
$   t_\mathrm{squeeze} \approx \SI{30}{\mu s}\times n_\mathrm{cycles} (2\pi\times\SI{50}{kHz}/\omega_{z}) 
$ for e.g. $\omega_i'/\omega_i = 0.2$.

\textit{Impulse sensing}--- At the end of the squeezing protocol, the mechanical state is momentum-squeezed, and will have a large associated delocalization in position. If the particle continues to evolve under the influence of the harmonic trapping potential, its Wigner function will rotate in phase space, thus reducing the useful momentum-squeezing. As such, once the particle has become maximally momentum-squeezed, one may turn off the trap and allow the particle to freely evolve, in order that the metrological advantage in impulse sensing of the squeezing is not lost.

One immediate consequence of turning off the trap is that the nanoparticle falls due to gravity. This can be dealt with in various ways. One may electrically charge the nanoparticle and apply an electric field to compensate for the gravitational force, as recently experimentally implemented~\cite{Steiner:2025dru}. Alternatively, one can simply let the particle fall and subsequently recapture it by turning the trap back on, which has also been demonstrated experimentally~\cite{Mattana:2025sre} (see also~\cite{Prakash:2025qvi} for a related implementation of levitated optomechanics in a drop-tower). The latter option suffers less from decoherence due to stray electric fields, so we proceed to analyze this case. 

In the absence of a trapping potential, the oscillator evolves according to Eq.~\eqref{eqn:SME} with $H_0 = p_i^2/2m$ and $\eta_i = 0$. Furthermore, the decoherence rates $\Gamma_i$ are now no longer dominated by the laser backaction, but by the emission of blackbody radiation from the thermal nanoparticle. In this case, the equations of motion for the covariances are analogous to Eq.~\eqref{eqn:cumulants}, and may be solved exactly. After undergoing free expansion for a time $\Delta t = t -t_0$, the covariances read
\begin{align}
\begin{split}
    V_{x_i x_i}(t) &= V_{x_i x_i}(t_0) + \frac{2\Delta t}{m} C_{x_i p_i}(t_0)  + \frac{\Delta t^2}{m^2} V_{p_i p_i}(t_0), \\
    V_{p_i p_i}(t) &= V_{p_i p_i}(t_0) + D_\mathrm{BB} \Delta t, \\
    C_{x_i p_i}(t) &= C_{x_i p_i}(t_0) + \frac{\Delta t}{m} V_{p_i p_i(t_0)} + \frac{D_\mathrm{BB}\Delta t^2}{2m} ,
    \end{split}
    \label{eqn:freeExpansion}
\end{align}
where $D_\mathrm{BB} = \Gamma_{\mathrm{BB},i}/x_{0,i}^2$ is the diffusion rate due to black body emission, given by \eqref{eqn:blackbodyDecoherence}. Eq.~\eqref{eqn:freeExpansion} encapsulates the three effects that occur during the free expansion stage of the protocol, in the absence of a signal. The first two are purely unitary effects: the particle becomes delocalized, with the position variance growing quadratically in time, and experiences a ``shear'' as its covariance grows linearly with time. The third is diffusion in momentum space due to blackbody radiation, which we find to be negligible. 

After the period of free evolution, the particle must be measured in order to infer any impulses that have acted. The putative presence of an impulse is rotated into the position quadrature by turning the potential $\omega$ back on for a quarter-period, at which point the position of the particle is projectively measured with a short laser pulse of length $t_\mathrm{meas}$. For such a projective measurement, the noise is dominated by the width of the state, and so the increase in SNR of the measurement is set by the degree of squeezing of the state that we previously characterized. This projective measurement requires both that the measurement rate is much larger than the natural frequency of the oscillator $\Gamma_i \gg \omega_i$ and that the shot noise-limited uncertainty on the measurement $\Delta x_i = x_{0,i}/\sqrt{\eta_i \Gamma_i t_\mathrm{meas}}$ is much smaller than the width of the state. The former requirement necessitates an unfocused laser beam, while the latter requires a large number of photons. With these requirements satisfied, the impulse sensing protocol can achieve a resolution more than 15 dB below the zero-point width with an appropriately timed $\SI{10}{\mu J}$ pulsed laser (see End Matter). 

For use in practice with signals at unknown times, an important consideration in the duration of the free-fall time is the duty cycle of the protocol. Once the projective measurement has been carried out, the particle is cooled back to its ground state, and the squeezing protocol begins again. Compared to a free-fall time of $\sim$0.5 ms, set by the requirement that the particle experience a linear trapping force~\cite{Mattana:2025sre}, the amount of time spent squeezing the particle is an order of magnitude smaller, as shown at the end of the previous section. Assuming the cooling can be achieved in similar time to the squeezing, most of the time is spent in the metrologically useful squeezed state. 

Technical noise must also be minimized in order to take advantage of this metrological protocol. Relevant sources are stochastic fluctuations in the laser intensity, which can decohere the particle~\cite{schneider1999decoherence, weiss2021large}, as well as vibrations between the control and measurement laser, which impact the precision of the final position measurement, similarly to hybrid RF-optical set-ups~\cite{bonvin2024hybrid}.

\textit{Outlook.}--- We have theoretically demonstrated a method of three-dimensional squeezing of the motional state of an optically levitated nanoparticle. The ultimate degree of squeezing is determined by the decoherence associated with the laser's recoil heating. For the parameters shown in Table~\ref{tab:params}, we predict that this enables approximately 15 dB (5 dB) of squeezing of the momentum variance of the particle's state in the direction perpendicular (parallel) to the laser's axis. This protocol is readily implementable with current technology, and could be useful for the next-generation of ultra-low threshold experiments in fundamental physics, such as searches for dark matter~\cite{Tseng:2025rlo} and sterile neutrinos~\cite{Carney:2022pku}. Tangentially, these protocols have also been considered as tools to accelerate the entanglement rate between two mechanical oscillators~\cite{Cosco:2020mhy}, for which our analysis may prove useful. The method may also be generalized to other trapping mechanisms, for instance involving dark traps, e.g., electric~\cite{bonvin2024hybrid} or optical~\cite{dago2024stabilizing}, with correspondingly lower decoherence rates, which may push these protocols even further below the SQL in the future. 

\begin{acknowledgments}
\textit{Acknowledgments.} The authors would like to thank Azriel Goldschmidt, Ryan Plestid and Madelyn Rahimi for helpful conversations. Work at Berkeley Lab is supported by the U.S. DOE, Office of High Energy Physics, under Contract No. DEAC02-05CH11231, the Quantum Horizons: QIS Research and Innovation for Nuclear Science Award DE-SC0023672, and an Early Career Research Award DE-SCL0000025. DCM is supported by in part, by NSF Grant PHY-2512192, the DOE Office of Science under Grants DE-SC0023672 and DE-SC0026367, and ONR Grant No. N00014-23-1-2600.
\end{acknowledgments}

\bibliographystyle{h-physrev}
\bibliography{references.bib}

\onecolumngrid
\section*{End Matter}
\twocolumngrid

\setcounter{equation}{0}
\renewcommand{\theequation}{A\arabic{equation}}

\begin{figure}
    \centering
    \includegraphics[width = \linewidth]{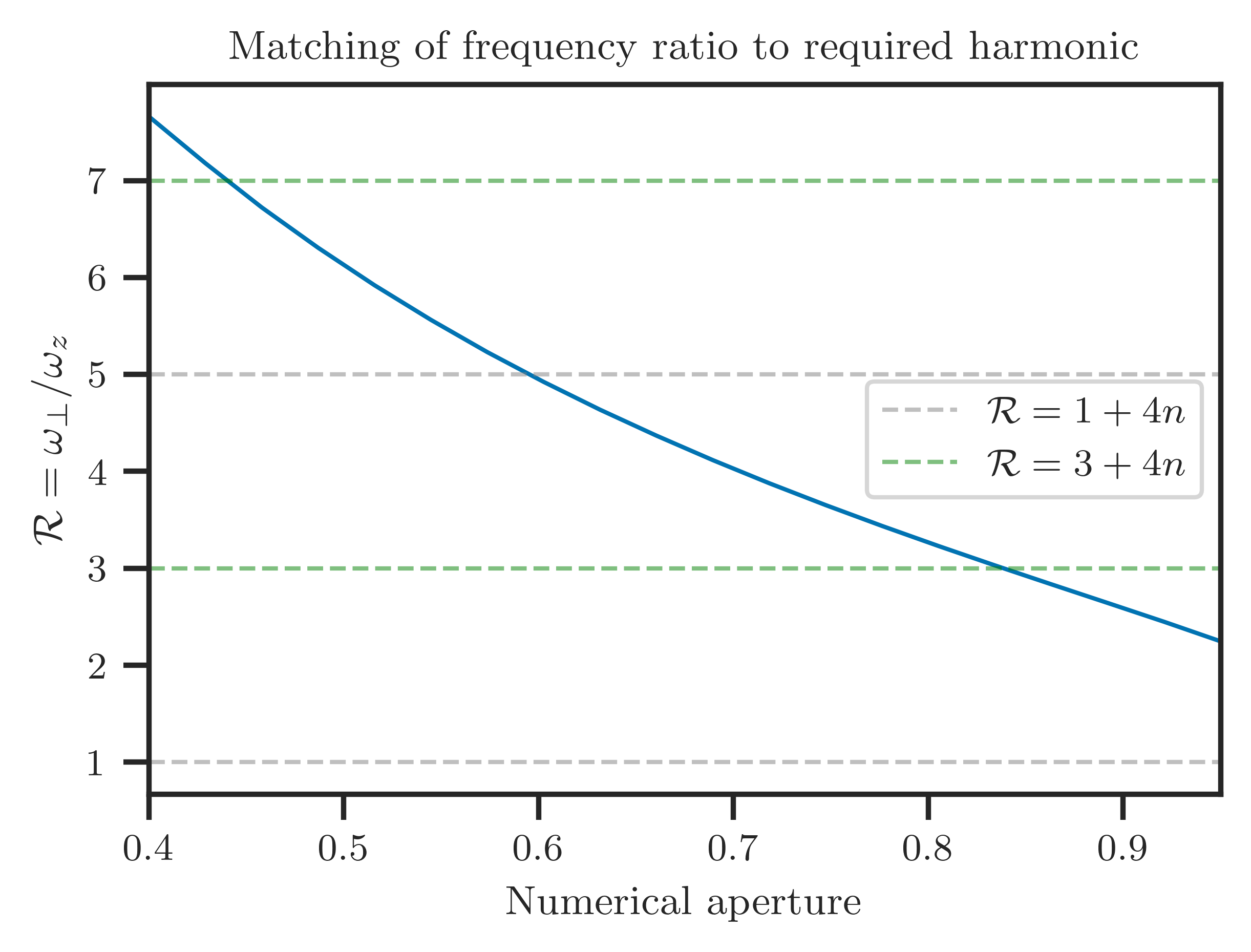}
    \caption{The ratio of the transverse to longitudinal frequency for a circularly polarized laser. The gray and green dashed lines correspond to the two different harmonic matching conditions of \eqref{eqn:harmonicCondition} with $k_z = 0$.}
    \label{fig:frequencyRatio}
\end{figure}

\textit{Optical potential}--- We now provide details of how the harmonic condition~\eqref{eqn:harmonicCondition} may be achieved in the field of a strongly focused tweezer. Throughout, we take $c = \hbar = \epsilon_0 = 1$, and we write the real electric field in terms of a complex vector $\mb{\tilde{E}}(t) = \re \mb{E} e^{-i\omega_\mathrm{L} t}$ where $\omega_\mathrm{L} = k_\mathrm{L} = 2\pi/\lambda$. The force on the dielectric particle, when averaged over many oscillation periods of the electric field, reads~\cite{novotny2012principles}
\begin{align}
 \mb{F}  = \frac{\alpha_1}{2} \re E_i \cdot \boldsymbol{\nabla} E_i^* + \frac{\alpha_2}{2} \im E_i \cdot \boldsymbol{\nabla} E_i^* ,
 \label{eqn:tweezerForce}
\end{align}
where $\alpha = \alpha_1 + i \alpha_2$ is the complex polarizability of the particle. Note that for a sub-wavelength, homogeneous sphere, we have a dynamic contribution to the imaginary component of the polarizability
$\alpha_2 = k_\mathrm{L}^3\alpha_1^2/6\pi$,
while $
\alpha_1 = 3 V_\mathrm{sph} (\epsilon_\mathrm{r}-1)/(\epsilon_\mathrm{r}+2)$
for an isotropic sphere of volume $V_\mathrm{sph}$. We numerically evaluate the force resulting from the electric field of a fundamental Gaussian beam focused by a spherical lens of arbitrary numerical aperture via the Richards-Wolf parametrization~\cite{richards1959electromagnetic}, from which we find the equilibrium position $\mb{x}_\mathrm{eq}$ of the nanoparticle. The spring constants in the three directions are given by the eigenvalues of the matrix 
$K_{ij} = -\nabla_j F_i(\mb{x}_\mathrm{eq})\label{eqn:springs}$, from which the natural frequencies follow. The ratio $\omega_\perp/\omega_z$ as a function of the numerical aperture is shown in Fig.~\ref{fig:frequencyRatio}, where we highlight the solutions to Eq.~\ref{eqn:harmonicCondition}.

\setcounter{equation}{0}
\renewcommand{\theequation}{B\arabic{equation}}

\textit{Decoherence rates}--- Here we overview the most relevant sources of decoherence, and how they depend on the experimental parameters. Consider time-evolution of the density matrix under a dissipator that induces decoherence in the position basis
\begin{align}
  d\rho &= -\frac{1}{2} \Gamma \frac{[x,[x,\rho]]}{x_0^2} dt,
\end{align}
where $x_0^2 = 1/(m \omega)$. In the presence of such a dissipator, an oscillator in a harmonic potential of frequency $\omega$ heats up at a rate $d \langle  a^\dag a \rangle/dt = \Gamma$. Hence we may identify the decoherence rate $\Gamma$ of an interaction with its heating rate, which facilitates the following semiclassical calculations.

First, we consider the decoherence induced by scattering of tweezer light off the sub-wavelength, dielectric sphere. For incident circularly polarized light, the probability for scattering a photon of momentum $\mb{k}$ into momentum $\mb{k}'$ with polar angle $\theta$ and azimuthal angle $\phi$ is 
$w(\theta, \phi) = 3\left(1 - \sin^2\theta\right)/4\pi,$
where we align the azimuthal axis with the longitudinal $z$-axis.
For elastic scattering $|\mb{k}'| = |\mb{k}| = k$, we can use this to find the angle-average of the squared momentum transfer $
\langle q_i^2 \rangle = k^2
 \int d\Omega \, (\hat{k}_i'^2 + \hat{k}_i^2 - 2\hat{k}_i' \hat{k}_i)  w(\Omega).
$
In terms of this squared momentum transfer, the rate of change of the $i$th mode's occupation number is 
\begin{align}
\Gamma_i = \Gamma_\mathrm{scatt} \frac{\langle q_i^2 \rangle_\lambda}{2m \omega_i},
\label{eqn:heatingRate1}
\end{align}
where $\omega_i$ is the angular frequency of the $i$th mode, and the scattering rate $\Gamma_\mathrm{scatt} $ is given in terms of the laser intensity at the sphere location $I_0$, the scattering cross-section $\sigma = \alpha_2 k$, and the laser frequency $\omega$, as
$\Gamma_\mathrm{scatt} = I_0 \alpha_2$.
Putting this together, we find the ratio
\begin{align}
\begin{split}
\frac{\Gamma_i}{\omega_i} &=   \frac{I_0}{2 K_i} \alpha_2 \langle q_i ^2 \rangle_\lambda 
\label{eqn:heatingRate2}
\end{split}
\end{align}
where $K_i = m \omega_i^2$ is the spring constant. This equation gives us the heating rate of the $i$th mode relative to its natural frequency in terms of calculable properties of the electric field, which we evaluate numerically in an analogous way to the trapping frequency. In the low NA regime, the analytic expression
\begin{align}
\begin{split}
    \Gamma_i &= \frac{3 \mathrm{NA}^2 P_\mathrm{L} \omega_\mathrm{L}^7 V_\mathrm{sph}c_i}{8 \pi^2 \rho \omega_i}\frac{(\epsilon_\mathrm{r}-1)^2}{(2 + \epsilon_\mathrm{r})^2} \\
    &\approx 2\pi\times\SI{1}{GHz} \;\frac{c_i}{c_z}\left(\frac{\mathrm{NA}}{0.25}\right)^2\frac{P_\mathrm{L}}{\SI{100}{W}} \left(\frac{\SI{532}{nm}}{\lambda}\right)^7,
    \label{eqn:lowNArecoil}
    \end{split}
\end{align}
may be used~\cite{gonzalez-ballesteroSuppressing2023}, where $c_i = (1,2,7)/5$.

An irreducible source of decoherence arises from the localizing effect of blackbody emission from the surface of the nanoparticle at internal temperature $T_\mathrm{int}$. The decoherence rate due to this blackbody emission dominates during the period of free-evolution, and is given by~\cite{romero2011quantum}
\begin{align}
\begin{split}
    \Gamma_\mathrm{BB} &= \frac{4\pi^4}{63}V x_0^2  T_\mathrm{int}^6 \mathrm{Im} \, \frac{\epsilon_\mathrm{BB} - 1}{\epsilon_\mathrm{BB} + 2} \\
    &\approx \SI{e-5}{Hz} \left(\frac{T_\mathrm{int}}{\SI{200}{K}}\right)^6,
    \label{eqn:blackbodyDecoherence}
    \end{split}
\end{align}
where we use the parameters of Tab.~\ref{tab:params}. Meanwhile, the average emission rate of these blackbody photons is~\cite{chang2010cavity}
\begin{align}
\begin{split}
   \gamma_\mathrm{BB} &= \frac{72 \zeta(5)}{\pi^2} V_\mathrm{sph} T_\mathrm{int}^4 \mathrm{Im} \, \frac{\epsilon_\mathrm{BB} - 1}{\epsilon_\mathrm{BB} + 2} \\
   &\approx \SI{30}{MHz} \left(\frac{T_\mathrm{int}}{\SI{200}{K}}\right)^4 ,
   \end{split}
\end{align}
where $\zeta(5) \approx 1.04$ is a value of the Riemann zeta function. This rate is much more frequent than the characteristic frequency of the mechanical motion, justifying the treatment of blackbody emission as Brownian diffusion in \eqref{eqn:freeExpansion}. 

Gas collisions may also decohere the nanoparticle. The average rate of collisions between the nanoparticle and a gas particle is $\gamma_\gas = n_\gas \langle \sigma v\rangle_\gas,$
where $n_\gas$ is the number density of gas molecules (assuming a single gas species), and $\langle \sigma v\rangle_\gas$ is the product of the cross-section and the relative velocity, averaged over the gas's phase space. For a gas particle following an isotropic, Maxwell-Boltzmann distribution at temperature $T$, geometric scattering off a spherical particle of radius $R_\mathrm{sph}$ gives
\begin{align}
    \langle \sigma v\rangle_\gas = \frac{\pi R_\sph^2}{4} \sqrt{ \frac{8T}{ \pi m_\gas } },
    \label{eqn:gasHardScatter}
\end{align}
for a gas particle of mass $m_\gas$, which we take to be that of molecular hydrogen. With the use of the ideal gas law, we then find
\begin{align}
    \gamma_\gas &= P R_\sph^2 \sqrt{\frac{\pi}{2 m_\gas T}}    \approx \SI{20}{Hz} \label{eqn:gasScatter}
\end{align}
which is much less than the mechanical response time of the oscillator, implying that an individual trajectory of the oscillator does not behave as though it is undergoing diffusion through Brownian motion. Moreover, most runs of the experimental protocol do not see a gas collision.

\setcounter{equation}{0}
\renewcommand{\theequation}{C\arabic{equation}}

\textit{Squeezing in absence of measurement}--- In the limit that the measurement efficiencies $\eta_i$ vanish, \eqref{eqn:cumulants} reduces to a set of coupled linear, ordinary differential equations, and can be solved exactly for the piecewise-constant potentials $\omega(t)$ that we consider here. Under evolution in a potential of frequency $\omega'$, one may solve the Heisenberg equations of motion for the position and momentum operators. Suppressing mode indices, we recognize the transformation
\begin{align}
    \begin{pmatrix}x(t) \\ \frac{p(t)}{m\omega}\end{pmatrix} = \begin{pmatrix}
        0 & \pm1 \\ \mp1 & 0 
    \end{pmatrix}\begin{pmatrix}
        \omega'/\omega & 0 \\
        0 & \omega/\omega'
    \end{pmatrix}
     \begin{pmatrix}x(t_0) \\ \frac{p(t_0)}{m\omega}\end{pmatrix}
\end{align}
as a composition of rotation and squeezing~\cite{janszky1992strong,weedbrook2012gaussian}, where the upper (lower) sign corresponds to evolution by a quarter-period (three-quarter-period). This implies the second cumulant vector $\mathbf{V} = (V_{xx}, V_{pp}, C_{xp})$ obeys a linear equation
\begin{align}
    \mathbf{V}(t) = M(t,t_0) \cdot \mb{V}(t_0) + \mb{D}(t,t_0),
\end{align}
where the matrix $M$ encodes the evolution induced by the time-varying harmonic potential, and the vector $\mb{D}$ encodes the additional diffusion due to decoherence.

We now consider a time-window $[t,t_0]$ within which the particle sits in a constant potential $\omega' < \omega$ with a constant heating $\Gamma(t) = \Gamma$ for a quarter-period $\pi /(2 \omega')$ , in which case
\begin{align}
    M(t-t_0) = \begin{pmatrix}
        0 & \frac{1}{m^2 \omega'^2} & 0 \\m^2 \omega'^2 & 0 & 0 \\
        0 & 0 & -1
    \end{pmatrix},
\end{align}
and the particle diffuses by an amount 
\begin{align}
    \mathbf{D} = \frac{\Gamma}{x_0^2 \omega'} \times \begin{pmatrix}
        \frac{\pi  }{m ^2\omega'^2} \\
        \pi \\
        \frac{1}{m \omega'}
    \end{pmatrix}.
\end{align}

In the case that the diffusion is dominated by the laser noise, the diffusion rate is proportional to the strength of the harmonic potential, and so we may write $\Gamma(t) = \Gamma \frac{\omega(t)}{\omega}$, where $\Gamma$ is the rate at the nominal trap frequency $\omega$. Assuming we start in a thermal state of a harmonic oscillator of frequency $\omega$ with occupation number $n_\mathrm{t}$, the momentum variance after this one pulse sequence is
\begin{align}
    V_{pp}^{(1)} = \frac{m \omega}{2 } \left( e^{-2\bar{r}}(1 + 2 n_\mathrm{t}) +   \frac{\pi}{2} \frac{\Gamma}{\omega} \right),
\end{align}
where we have defined 
$    \bar{r} = \ln{\frac{\omega}{\omega'}}
   $
as the squeezing strength in a single pulse sequence in the absence of decoherence. 

We next go back to a potential at $\omega$ for a quarter-period, before returning to $\omega'$ again. One can show that the momentum variance after $N_\mathrm{p}$ such pulse cycles is
\begin{align}
\begin{split}
    V_{pp}^{(N_\mathrm{p})} = &\frac{m \omega}{2 } \Big[ e^{-2N_\mathrm{p}\bar{r}}(1 + 2 n_\mathrm{t}) \\
    &+   \frac{\pi}{2} \frac{\Gamma}{\omega}\left(-1 + 2\sum_{k=0}^{N_\mathrm{p}}e^{-2N_\mathrm{p}\bar{r}} \right) \Big].
    \label{eqn:varianceFiniteNc}
    \end{split}
\end{align}
In the limit of a large number of cycles, the terms on the first line of \eqref{eqn:varianceFiniteNc} vanish, and we may resum the second line to find the result given in Eq.~\eqref{eqn:asymptoticSqueezing} in the case that the initial state is a coherent state.

\end{document}